\documentclass[prb,superscriptaddress,floatfix,twocolumn]{revtex4}
\usepackage{amsfonts}
\usepackage{stmaryrd}
\usepackage{bbm}
\usepackage{mathrsfs}
\usepackage{tipa}
\usepackage{amssymb}
\usepackage{txfonts}
\usepackage{graphicx}
\usepackage{dcolumn}
\usepackage{epstopdf}
\usepackage{array}
\usepackage{color}

\newcommand{\PreserveBackslash}[1]{\let\temp=\\#1\let\\=\temp}
\newcolumntype{C}[1]{>{\PreserveBackslash\centering}p{#1}}
\newcolumntype{R}[1]{>{\PreserveBackslash\raggedleft}p{#1}}
\newcolumntype{L}[1]{>{\PreserveBackslash\raggedright}p{#1}}

\begin{document}

\newcommand*{\cm}{cm$^{-1}$\,}

\title{Electronic structure and open-orbit Fermi surface topology in isostructural semimetals NbAs$_2$ and W$_2$As$_3$ with extremely large magnetoresistance}

\author{Rui Lou$^{1,\,2,\,3,\,4,\,\rm a)}$}

\author{Yiyan Wang$^{2,\,5}$}

\author{Lingxiao Zhao$^{6,\,7}$}

\author{Chenchao Xu$^{8}$}

\author{Man Li$^{2,\,9}$}

\author{Xiaoyang Chen$^{3}$}

\author{Anmin Zhang$^{1,\,2}$}

\author{Yaobo Huang$^{9}$}

\author{Chao Cao$^{8}$}

\author{Genfu Chen$^{10,\,11}$}

\author{Tianlong Xia$^{2,\,\rm a)}$}

\author{Qingming Zhang$^{1,\,10}$}

\author{Hong Ding$^{10,\,11,\,12}$}

\author{Shancai Wang$^{2,\,\rm a)}$}

\begin{abstract}
  \vspace{-1.4em}
  \setlength{\baselineskip}{11pt}
  \emph{\\$^1$School of Physical Science and Technology, Lanzhou University, Lanzhou 730000, China
  \\$^2$Department of Physics and Beijing Key Laboratory of Opto-electronic Functional Materials $\textsl{\&}$ Micro-nano Devices, Renmin University of China, Beijing 100872, China
  \\$^3$State Key Laboratory of Surface Physics, Department of Physics, and Laboratory of Advanced Materials, Fudan University, Shanghai 200438, China
  \\$^4$Leibniz Institute for Solid State and Materials Research, IFW Dresden, 01069 Dresden, Germany
  \\$^5$Institute of Physical Science and Information Technology, Anhui University, Hefei 230601, China
  \\$^6$Wuhan National High Magnetic Field Center, Huazhong University of Science and Technology, Wuhan 430074, China
  \\$^7$School of Physics, Huazhong University of Science and Technology, Wuhan 430074, China
  \\$^8$Department of Physics, Zhejiang University, Hangzhou 310027, China
  \\$^9$Shanghai Synchrotron Radiation Facility, Shanghai Institute of Applied Physics, Chinese Academy of Sciences, Shanghai 201204, China
  \\$^{10}$Beijing National Laboratory for Condensed Matter Physics, and Institute of Physics, Chinese Academy of Sciences, Beijing 100190, China
  \\$^{11}$Songshan Lake Materials Laboratory, Dongguan, Guangdong 523808, China
  \\$^{12}$CAS Center for Excellence in Topological Quantum Computation, University of Chinese Academy of Sciences, Beijing 100049, China}
  \\\textcolor{black}{$^{\rm a)}$Authors to whom correspondence should be addressed: Rui Lou, lourui@lzu.edu.cn; Tianlong Xia, tlxia@ruc.edu.cn;
  and Shancai Wang, scw@ruc.edu.cn}
  \setlength{\parskip}{1.1\baselineskip}

  In transition-metal dipnictides $TmPn_2$ ($Tm$ = Ta, Nb; $Pn$ = P, As, Sb), the origin of extremely large magnetoresistance (XMR) is yet to be
  studied by the direct visualization of the experimental band structures. Here, using angle-resolved photoemission spectroscopy, we map out the three-dimensional electronic structure of NbAs$_2$. The open-orbit topology contributes to a non-negligible part of the Fermi surfaces (FSs),
  like that of the isostructural compound MoAs$_2$, where the open FS is proposed to likely explain the origin of XMR. We further demonstrate the
  observation of open characters in the overall FSs of W$_2$As$_3$, which is also a XMR semimetal with the same space group of $C$12/$m$1 as $TmPn_2$
  family and MoAs$_2$.
  Our results suggest that the open-orbit FS topology may be a shared feature between XMR materials with the space group of $C$12/$m$1, and thus could
  possibly play a role in determining the corresponding XMR effect together with the electron-hole compensation.
\end{abstract}

\maketitle

Serving as a fertile ground for topological quantum states and a promising candidate for device applications, the topological semimetal (TSM) has
inspired great research interest in the community \cite{HMWeng2014,HMWeng2016,XGWan2011,ZJWang2012,HWWeng2015,SMHuang2015,AASoluyanov2015,
GChang2018,ZKLiu2014,SBorisenko2014,BQLv2015X,SXu2015S,LXYang2015,SXu2015NP,ZZhu2016,BQLv2017,QXu2015,RLou2016,LMSchoop2016,XZhang2017,XFeng2018,
RLou2018X,RLou2018npj}. Different from topological insulators \cite{MZHasan2010}, the bulk states and topological surface states in TSMs
have more novel forms \cite{BQLv2015X,SXu2015S,LXYang2015,SXu2015NP,SXu2015arc,RLou2018X,RLou2018npj,GBian2016NC,ZKLiu2014,SBorisenko2014,BQLv2015NP}.
With such diverse electronic properties, TSMs exhibit many unusual magnetotransport phenomena, such as linear transverse magnetoresistance
(MR) and negative longitudinal MR in Dirac/Weyl semimetals \cite{JFeng2015,TLiang2015,XCHuang2015,CShekhar2015,JXiong2015,CZhang2016NC,CZLi2015} and
extremely large transverse MR (XMR) in nonmagnetic semimetals \cite{LSchubnikow1930,PBAlers1951,TKasuya1993,FYYang1999,RXu1997,EMun2012}. Fully exploring
the MR-related phenomena can not only facilitate understanding the physics in TSMs, but also promote the development of practical applications like
magnetic memory/sensor devices.

\begin{figure*}[t]
  \setlength{\abovecaptionskip}{-0.4cm}
  \setlength{\belowcaptionskip}{-0.05cm}
  \begin{center}
  \includegraphics[trim = 0mm 0mm 0mm 0mm, clip=true, width=1.85\columnwidth]{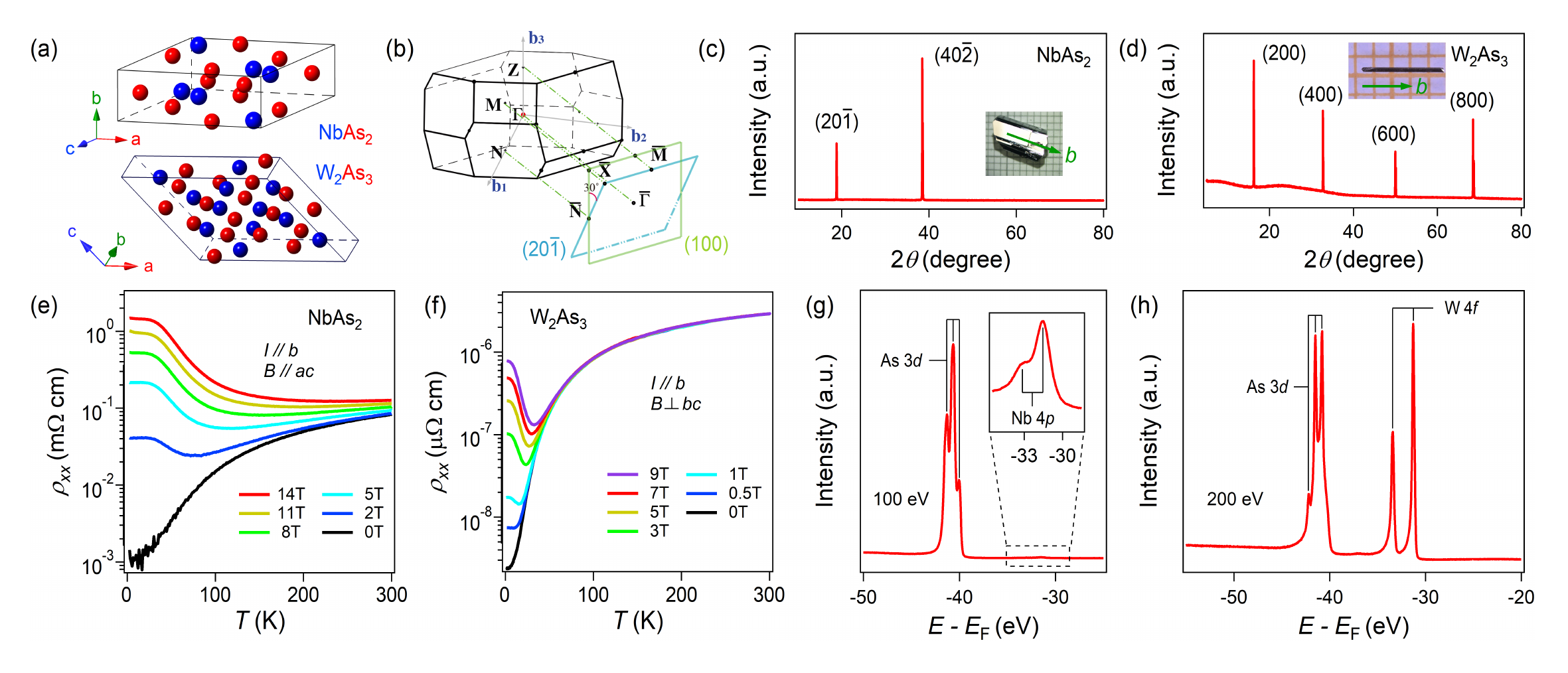}
  \end{center}
  \caption{Single crystals of NbAs$_2$ and W$_2$As$_3$.
  (a) Schematic crystal structures of NbAs$_2$ and W$_2$As$_3$.
  \textcolor{black}{(b) Sketches of the 3D primitive BZ and conventional BZs of (20$\overline{1}$) and (100) surfaces of monoclinic $C$12/$m$1 space group structure.}
  (c) XRD pattern on the NbAs$_2$ (20$\overline{1}$) surface. Inset: Picture of a NbAs$_2$ crystal.
  (d) Same as (c) on the W$_2$As$_3$ (200) surface. Inset: Picture of a W$_2$As$_3$ crystal.
  (e) Temperature dependence of the resistivity ($\rho_{xx}$) of NbAs$_2$ under different magnetic fields, where $I$ is parallel to the $b$ axis
  and $B$ is parallel to the $ac$ plane.
  (f) Same as (e) of W$_2$As$_3$, where $I$ is parallel to the $b$ axis and $B$ is perpendicular to the $bc$ plane.
  (g),(h) Core-level photoemission spectra of NbAs$_2$ and W$_2$As$_3$ measured with 100- and 200-eV photons, respectively.
  }
\end{figure*}

Recently, since the discovery of quadratic XMR effect in WTe$_2$ \cite{MNAli2014}, significant research interest has been stimulated in understanding its microscopic mechanism and exploring new XMR materials. A series of nonmagnetic semimetals with XMR have been established, like transition-metal dipnictides $TmPn_2$ ($Tm$ = Ta, Nb; $Pn$ = P, As, Sb)\textcolor{black}{\cite{KWang2014,BShen2016,DSWu2016,CCXu2016,YYWang2016,ZWangarXiv,SJia2016,
YLuo2016TA2,YPLi2016arXiv,APariari2018TaSb2,KYokoi2018PRM,GPeramaiyan2018,
Leahy2018PNAS,LGuo2018JAP,VHarimohan2019,XRao2019arXiv,TAButcher2019,HYWang2019PRB,YKLi2016TaSb2,PKSudesh2019,XLiu2020rev,SLee2021PNAS}}, $LnX$ ($Ln$ = La, Y, Nd, or Ce; $X$ = Sb/Bi) \cite{FFTafti2015,SSSun2016,NKumar2016,QYu2017,OPavlosiuk2016,NWakeham2016,NAlidoustarXiv,XDuan2018,CGuo2017}, ZrSiS family \cite{RSingha2017,MNAli2016,XWang2016}, MoAs$_2$ \cite{RLou2017,JWang2017,RSingha2018}, W$_2$As$_3$ \cite{YPLi2018,LXZhao2019,JWang2019},
PtBi$_2$ \cite{WSGao2017}, PtSn$_4$ \cite{EMun2012}, WP$_2$ \cite{AWang2017}, etc.
So far, several scenarios have been proposed to explain the XMR behavior, including nontrivial band topology \cite{FFTafti2015}, electron-hole compensation \cite{MNAli2014,LKZeng2016}, open-orbit Fermi surface (FS) topology \cite{RLou2017}, forbidden backscattering at zero field \cite{JJiang2015}, and field-induced FS changes \cite{KWang2014}. However, when looking back one finds that the study of $TmPn_2$ family is still at the early stage it was while it ignited substantial works. The lack of experimental electronic structure study is the main problem because the MR behavior in semimetals is believed to intimately correlate with the underlying electronic structure.
\textcolor{black}{Previous angle-resolved photoemission spectroscopy (ARPES) \cite{RLou2017} and quantum oscillations \cite{RSingha2018} measurements on the isostructural compound MoAs$_2$ suggest that the origin of XMR may be attributed to the dominant open-orbit FS topology.}
In order to gain insights into the mechanism of
the XMR in these isostructural materials, it is critical to experimentally reveal the band structure of other representative compounds.

Utilizing ARPES, here we present the electronic structure study on XMR semimetals NbAs$_2$ and W$_2$As$_3$, both with the space group of $C$12/$m$1 \cite{YYWang2016,LXZhao2019}. Unlike most XMR materials whose FSs are usually dominated by closed pockets, we uncover that the band structures of NbAs$_2$
and W$_2$As$_3$ host non-negligible open-orbit FS topologies in the three-dimensional (3D) momentum space, where the ``ripple"-shaped bulk FSs show open character
along two perpendicular high-symmetry directions. Our results, combining with the previously studied MoAs$_2$ \cite{RLou2017,RSingha2018}, suggest that the open
FS may be a shared feature between XMR compounds with the space group of $C$12/$m$1, and thus could possibly make a contribution to the corresponding XMR behavior
together with the electron-hole compensation.

High-quality single crystals of NbAs$_2$ and W$_2$As$_3$
were obtained by chemical vapor transport methods \cite{YYWang2016,LXZhao2019}. ARPES measurements of NbAs$_2$ and W$_2$As$_3$
were performed at the Dreamline beamline of the Shanghai Synchrotron Radiation Facility and the beamline 5-2 of the Stanford Synchrotron
Radiation Lightsource (SSRL) using photons with linear horizontal polarization, respectively. The energy and angular resolutions were set to better than 15 meV
and 0.1$^{\circ}$, respectively.
Samples were cleaved \emph{in situ}, yielding flat mirrorlike surfaces. During the measurements, the temperature was kept at 20 K and the pressure was
maintained better than 5 $\times$ 10$^{-11}$ Torr.
The first-principles calculations were performed with Vienna ab initio Simulation Package \cite{VASP_Kresse_PRB93,VASP_Kresse_PRB96} for W$_2$As$_3$.
A plane-wave basis up to 400 eV was employed in the calculation. The Perdew-Bruke-Ernzerhof parametrization of generalized gradient approximation to the exchange correlation functional \cite{GGA_Perdew_PRL96} was employed in the calculation. The electronic structure was calculated for the conventional cell with 4 $\times$ 12 $\times$ 5 $\Gamma$-centered $k$ mesh. The tight-binding (TB) Hamiltonian was obtained by fitting the density-functional theory band structures using
maximally localized Wannier function method \cite{MLWF}. The electronic structure of (100) surface was calculated for the conventional cell with surface Green's function method \cite{Sancho_1985} using the TB Hamiltonian after symmetrization \cite{ZHI2022108196}.

\begin{figure}[t]
  \setlength{\abovecaptionskip}{-0.4cm}
  \setlength{\belowcaptionskip}{-0.1cm}
  \begin{center}
  \includegraphics[trim = 4.7mm 0mm 0mm 0mm, clip=true, width=1.05\columnwidth]{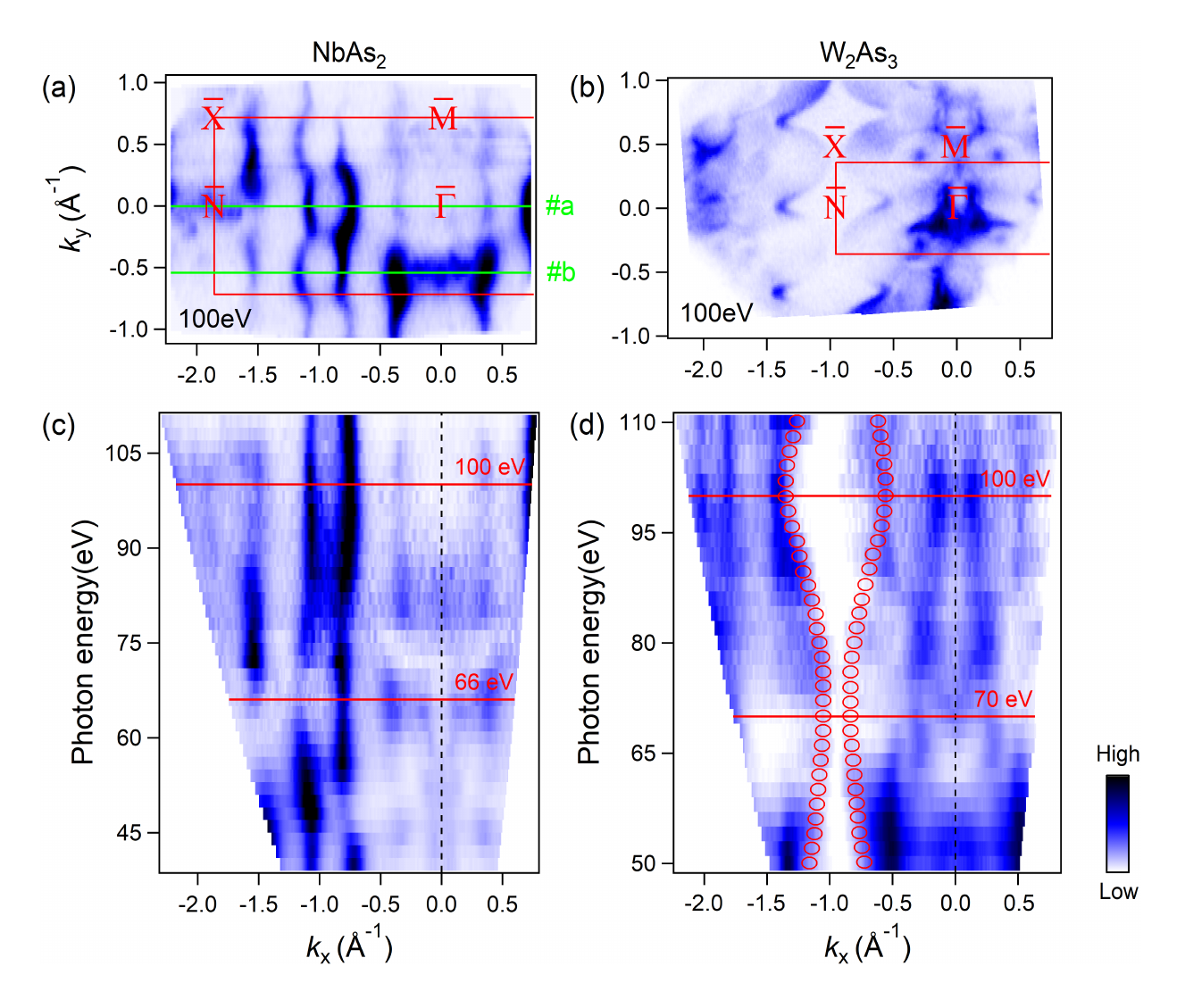}
  \end{center}
  \caption{3D FS topologies of NbAs$_2$ and W$_2$As$_3$.
  (a) Constant-energy ARPES image of NbAs$_2$ by integrating the photoemission intensity within $E_{\rm F}$ $\pm$ 40 meV measured with 100-eV photons.
  Cuts \#a and \#b indicate the locations of the experimental band structures in Fig. 3.
  (b) Same as (a) of W$_2$As$_3$. The red solid curves show the (20$\overline{1}$)- and (100)-projected BZs of NbAs$_2$ and W$_2$As$_3$, respectively.
  (c) ARPES intensity plot of NbAs$_2$ in the $h\nu$-$k_x$ plane at $E_{\rm F}$, with $k_x$ oriented along the \textcolor{black}{$\bar{\Gamma}$-$\bar{N}$} direction.
  (d) Same as (c) of W$_2$As$_3$. The modulation of open FS is indicated by red open circles.
  }
\end{figure}

\begin{figure*}[t]
  \setlength{\abovecaptionskip}{-0.3cm}
  \setlength{\belowcaptionskip}{-0.0cm}
  \begin{center}
  \includegraphics[trim = 0mm 0mm 0mm 0mm, clip=true, width=1.5\columnwidth]{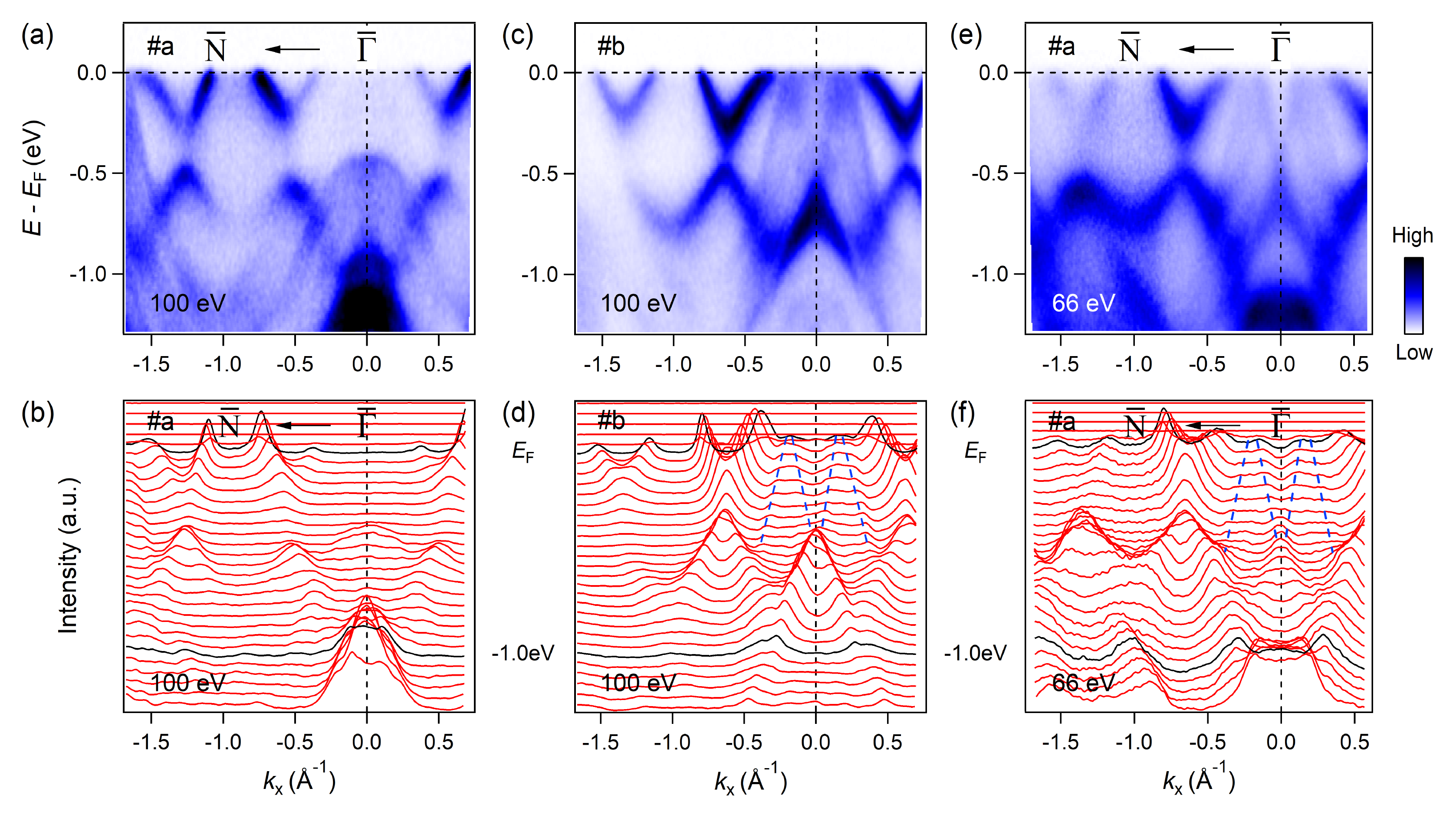}
  \end{center}
  \caption{Near-$E_{\rm F}$ band structure of NbAs$_2$.
  (a),(b) ARPES intensity plot and corresponding MDCs along the \textcolor{black}{$\bar{\Gamma}$-$\bar{N}$} direction [cut \#a in Fig. 2(a)] recorded at $h\nu$ = 100 eV, respectively.
  (c),(d) Same as (a),(b) along cut \#b in Fig. 2(a).
  (e),(f) Same as (a),(b) taken with 66-eV photons.
  The blue dashed lines in (d),(f) are extracted peak positions, indicating the linearly dispersing bands around $k_x$ $\sim$ 0.
  }
\end{figure*}

Figure 1 illustrates the overall physical properties of NbAs$_2$ and W$_2$As$_3$ crystals. The crystal structures with the monoclinic $C$12/$m$1 space
group are presented in Fig. 1(a).
\textcolor{black}{The corresponding bulk primitive Brillouin zone (BZ) and conventional BZs of (20$\overline{1}$) and (100) surfaces are shown in Fig. 1(b).}
Single crystal x-ray diffraction (XRD) patterns in
Figs. 1(c) and 1(d) demonstrate the crystal surfaces of our measured samples, that the (20$\overline{1}$) plane of NbAs$_2$ and the (200) plane of W$_2$As$_3$. The temperature dependence of resistivity under various magnetic fields [Figs. 1(e) and 1(f)] agrees with the previous studies on $TmPn_2$ and W$_2$As$_3$ \cite{BShen2016,DSWu2016,SJia2016,YPLi2018,JWang2019}. As most nonmagnetic XMR semimetals \cite{FFTafti2015,SSSun2016}, the temperature-dependent $\rho_{xx}$ exhibits a typical metallic behavior under zero field, and upon turning on the magnetic field, the resistivity prominently increases and forms
a plateau at low temperatures ultimately. The quadratic MR of 1.0 $\times$ 10$^5$\% and 6.9 $\times$ 10$^4$\% with the magnetic field up to 14 T can be
found in our earlier transport studies on NbAs$_2$ \cite{YYWang2016} and W$_2$As$_3$ \cite{LXZhao2019}, respectively.
The characteristic peaks of Nb, W, and As elements are resolved in the core-level photoemission spectra
in Figs. 1(g) and 1(h), further verifying the chemical compositions of NbAs$_2$ and W$_2$As$_3$ samples. It is noted that the Nb-4$p$ peaks are rather
weak in Fig. 1(g), which could be due to the dramatic suppression of the photoionization cross sections of Nb 4$p$ in this photon energy regime as compared with
W 4$f$ and As 3$d$.

In earlier theoretical calculations \cite{CCXu2016,YPLi2018}, the overall bulk band structures of NbAs$_2$ and W$_2$As$_3$ indicate the coexistence
of electron- and hole-like bands crossing the Fermi level ($E_{\rm F}$). This appears to support the carrier compensation mechanism suggested by the
transport measurements based on the classic two-band model \cite{SJia2016,YPLi2018}. To gain more insights into the origin of the XMR behavior  \cite{YYWang2016,LXZhao2019}, we
systematically investigate their electronic structures by ARPES. As shown in Figs. 2(a) and 2(b), one can see that the in-plane FSs of both
compounds exhibit periodic modulations along the $k_x$ and $k_y$ directions. Notably, the ``ripple"-shaped FS contours elongated along the $k_y$
directions show no signature of closure, behaving as the open FSs, in sharp contrast to the complicated closed ones along the $k_x$ = 0 lines ($\bar{\Gamma}$-$\bar{M}$). We further carry out photon-energy-dependent measurements along the \textcolor{black}{$\bar{\Gamma}$-$\bar{N}$} directions. In Figs. 2(c) and 2(d), the noticeable dispersions of overall band structure as a function of photon energy
demonstrate the bulk origin of our observations. Moreover, one obtains that the ``ripple"-shaped FSs do not close along the $k_z$ directions either.
The modulation behavior of the open FSs in each compound is found to be equivalent along the $k_y$ and $k_z$ directions. The presence of open characters
in 3D FS topologies of NbAs$_2$ and W$_2$As$_3$ is similar to the isostructural MoAs$_2$ \cite{RLou2017,RSingha2018}, implying that the open FS may be a shared feature between XMR materials with the space group of $C$12/$m$1.

\begin{figure*}[t]
  \setlength{\abovecaptionskip}{-0.45cm}
  \setlength{\belowcaptionskip}{-0.0cm}
  \begin{center}
  \includegraphics[trim = 0mm 0mm 0mm 0mm, clip=true, width=2\columnwidth]{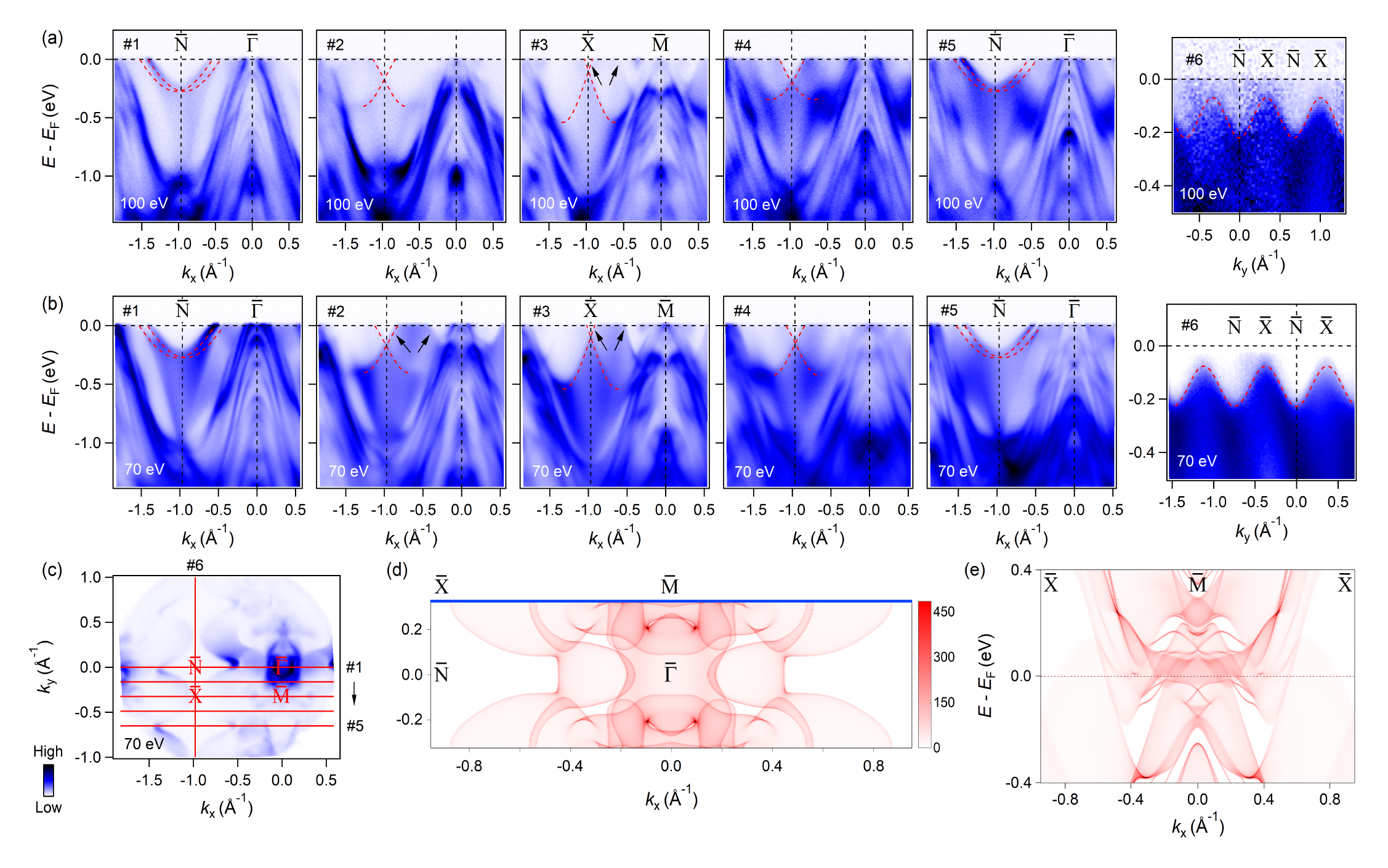}
  \end{center}
  \caption{Near-$E_{\rm F}$ band structure of W$_2$As$_3$.
  (a) ARPES intensity plots along cuts \#1-\#6 in (c) taken with 100-eV photons.
  (b) Same as (a) recorded at $h\nu$ = 70 eV. The red dashed curves in cuts \#1-\#5 are guides to the eye for the band crossings around the BZ boundary
  \textcolor{black}{($\bar{N}$-$\bar{X}$-$\bar{N}$)}.
  The black arrows indicate the band dispersions defining the open FSs. The red dashed curves in cuts \#6 are guides to the eye for the ``nodal lines".
  (c) Constant-energy ARPES image of W$_2$As$_3$ by integrating the photoemission intensity within $E_{\rm F}$ $\pm$ 40 meV measured with 70-eV photons.
  Cuts \#1-\#6 indicate the locations of the experimental band structures in (a),(b).
  (d) Calculated FS contours of the (100) surface of W$_2$As$_3$.
  (e) Calculated band structure along the $\bar{X}$-$\bar{M}$-$\bar{X}$ direction [blue solid line in (d)] of the (100) surface.
  }
\end{figure*}

The near-$E_{\rm F}$ ARPES spectra of NbAs$_2$ are measured along cuts \#a and \#b in Fig. 2(a). As shown in Figs. 3(a) and 3(c), the conduction band region
is composed of alternate electron-like bands, which define the neighbouring open FS sheets, and linearly dispersive hole-like bands corresponding to the closed pockets along the $\bar{\Gamma}$-$\bar{M}$ direction. By tracing the peak positions of momentum distribution curves (MDCs) in Figs. 3(b) and 3(d), one can observe the modulation of multiple linear dispersions (around $k_x$ $\sim$ 0) along the $k_y$ direction. We further record the band structure along cut \#a at $h\nu$ = 66 eV. The overall band dispersions [Figs. 3(e) and 3(f)] are analogous to that along cut \#b taken with 100-eV photons [Figs. 3(c) and 3(d)]. This similarity demonstrates that, including the open FSs, the modulation behavior of the electronic structure in NbAs$_2$ is equivalent along the $k_y$ and $k_z$ directions. Regardless of the FS topology, the observation of both electron- and hole-like bulk bands near $E_{\rm F}$ seems to favor the electron-hole compensation picture proposed in theoretical calculations \cite{CCXu2016} and transport studies \cite{SJia2016}, which will be discussed later.
\textcolor{black}{It is worth noting that the recent optical spectroscopy experiments on NbAs$_2$ have demonstrated the electromagnetic signature of dispersive Dirac nodal lines with spin-orbit coupling gaps \cite{YMShao2019,JWyzula2021}. The corresponding fermiology and band dispersions show similarities to that of the open FSs observed here. More unique properties associated with the open orbits thus merit further explorations.}

We then turn to study the band structure of W$_2$As$_3$. As illustrated in Fig. 2(d), the photon energies of 100 and 70 eV corresponding to the high-symmetry $k_z$ planes are used for the measurements. Because of the same modulation behavior of the open FSs along the $k_y$ and $k_z$ directions, the
neighbouring open FSs (in the $k_x$-$k_y$ plane), for example, at $k_y$ = 0 \textcolor{black}{($\bar{\Gamma}$-$\bar{N}$)}, are concave and convex with respect to the BZ boundary
\textcolor{black}{($\bar{N}$-$\bar{X}$)} in the 100- and 70-eV FS mappings, respectively. Here in Fig. 4(c), we artificially align the concave section of 70-eV mapping along
$k_y$ = 0 \textcolor{black}{($\bar{\Gamma}$-$\bar{N}$)} as in Fig. 2(b)
to unify the momentum cut assignment. As indicated by cuts \#1-\#5 in Fig. 4(c), we record the ARPES spectra near $E_{\rm F}$ along some representative
momentum cuts on the ``ripple"-shaped FSs. The intensity plots measured at $h\nu$ = 100 and 70 eV are shown in Figs. 4(a) and 4(b), respectively. The band structures around $\bar{\Gamma}$, $\bar{M}$, and along $k_x$ = 0 ($\bar{\Gamma}$-$\bar{M}$) are composed of multiple hole-like bands, forming complicated FS pockets [Figs. 2(b) and 4(c)].
In contrast, as highlighted by black arrows in Figs. 4(a) (cut \#3) and 4(b) (cuts \#2 and \#3), the relatively simple band structures, which define the ``ripple"-shaped open FSs, dominate around the BZ boundary \textcolor{black}{($\bar{N}$-$\bar{X}$)}. Moreover, from cuts \#1 to \#5, a band crossing formed by the dispersions of neighbouring
open FSs always exists. Accordingly, as illustrated in the spectra along cuts \#6 [indicated in Fig. 4(c)] in Figs. 4(a) and 4(b), the ``nodal-line"-like behavior is observed
in both two $k_z$ planes. As a result, these band crossings seem to constitute a ``nodal surface" normal to the $k_x$-$k_y$ plane. As suggested by an earlier
theoretical study on the nodal-surface semimetals, one necessary requirement for a topologically protected nodal surface is that the space-time inversion
symmetry is violated \cite{WWu2018}. However, the space inversion symmetry $P$ preserves in the space group of $C$12/$m$1 \cite{RLou2017} and there is
no magnetic order revealed in W$_2$As$_3$ \cite{YPLi2018,LXZhao2019,JWang2019}, thus the $P{\cdot}T$ symmetry is present. The ``nodal surface" here is
not topological and could be caused by band structure effects, such as band folding effect, which calls for future investigations. We further carry out
the TB calculations on the (100) surface of W$_2$As$_3$ using surface Green's function method. As shown in Figs. 4(d) and 4(e), one can obtain that the calculations within the first BZ well reproduce the overall electronic structures in experiment, especially the ones associated with the open characters.

The previous transport measurements have suggested the critical role of electron-hole compensation in determining the quadratic XMR behavior in NbAs$_2$ and W$_2$As$_3$ \cite{SJia2016,YPLi2018,LXZhao2019}. Although the carrier compensation scenario can be well applied in many XMR semimetals with closed FS trajectories
based on the two-band model \cite{SSSun2016,LKZeng2016,MNAli2014,ABPippard1989}, its availability in XMR materials with non-negligible open FS characters
is yet to be established \cite{NWAshcroft1976}. In materials with closed FSs, upon applying the magnetic field, all electron and hole orbits under Lorentz force are closed ones. The MR tends to saturate at high magnetic fields unless the electron and hole carriers are compensated, the dominant closed character of FSs is necessary for the XMR related to the carrier compensation mechanism.
In contrast, when the open orbits dominate the FS topology perpendicular to the magnetic field, the MR would increase as $H^2$ without any sign of saturation, regardless of whether or not the balance between electron and hole carriers \cite{JMZiman1972}. The distinct velocities of electrons in the plane perpendicular to the magnetic field are found to be the cause of these different behaviors \cite{AAAbrikosov1988,JSingleton2001}. Recently, the open-orbit FS topology has been suggested as a possible self-contained and dominant mechanism of the XMR in isostructural compound MoAs$_2$ \cite{RLou2017}.
Nevertheless, by comparing with the case in MoAs$_2$, where the open FS is reported to be dominant of the overall electronic structures \cite{RLou2017,RSingha2018}, our observations in NbAs$_2$ and W$_2$As$_3$ reveal that the open and closed features each defines a non-negligible part of FSs. Therefore, only considering the electron-hole compensation is inadequate, a collaboration between the open FS and the carrier compensation may explain the origin of the quadratic XMR in these two compounds. The open FS scenario may be taken into account in a broad range of XMR semimetals with the space group of $C$12/$m$1.

In summary, we have studied the electronic structure of XMR semimetals NbAs$_2$ and W$_2$As$_3$ by ARPES experiments. The ``ripple"-shaped bulk FSs do not
close along two perpendicular high-symmetry directions, showing the open-orbit FS topologies similar to that of isostructural MoAs$_2$ \cite{RLou2017,RSingha2018}.
Our results suggest that the open FS may be a shared feature between XMR materials with the space group of $C$12/$m$1, and thus could possibly collaborate
with the electron-hole compensation to solve the origin of the corresponding XMR effect.

The authors acknowledge Donglai Feng for very helpful discussions. In addition, we acknowledge Makoto Hashimoto and Donghui Lu for assisting with the
ARPES measurements at SSRL.
This work was supported by the National Natural Science Foundation of China (Grants No. 11904144,
No. 11774421, No. 11874422, No. 11874137, and No. 11704074) and the Chinese Academy of Sciences (CAS) (Project No. XDB07000000).
Y. B. H. was supported by the CAS Pioneer Hundred Talents Program.


\noindent\textbf{AUTHOR DECLARATIONS}

\noindent\textbf{Conflict of Interest}

The authors have no conflicts to disclose.

\noindent\textbf{Author Contributions}

R. L., Y. Y. W., L. X. Z., C. C. X., and M. L. contributed equally to this work.

\noindent\textbf{DATA AVAILABILITY}

The data that support the findings of this study are available from the corresponding authors upon reasonable request.

\end{document}